\documentclass[aps,prb,twocolumn,floatfix,showpacs,tightenlines,superscriptaddress]{revtex4-2}
\usepackage{mathtools}
\usepackage{times}
\usepackage{bm}
\usepackage{dsfont,amsthm,amsbsy}
\usepackage{verbatim}
\usepackage{amssymb}
\usepackage{amsmath}
\usepackage{bbm}
\usepackage{graphicx}
\usepackage{dcolumn}
\usepackage{epstopdf}
\usepackage{subfigure}
\usepackage{natbib}
\usepackage{epsfig}
\usepackage{amsfonts}
\usepackage{mathrsfs}
\usepackage{sidecap}
\usepackage{lipsum}
\usepackage[toc,page,title,titletoc,header]{appendix}
\usepackage[colorlinks,linkcolor=blue,citecolor=blue,anchorcolor=blue,urlcolor=blue]{hyperref}
\usepackage{hyperref}
\usepackage{resizegather}
\usepackage{tikz}
\usepackage{float}
\usepackage{mathbbol}
\usepackage[normalem]{ulem}
\usepackage{cancel}
\usepackage{upgreek}
\newcommand{\bl}{\begin{aligned}}
\newcommand{\el}{\end{aligned}}
\def\be{\begin{equation}}
\def\ee{\end{equation}}

\def\bi{\begin{itemize}}
\def\ei{\end{itemize}}
\def\bn{\begin{enumerate}}
\def\en{\end{enumerate}}
\def\bea{\begin{eqnarray}}
\def\eea{\end{eqnarray}}
\def\no{\nonumber}
\def\ba{\begin{array}}
\def\ea{\end{array}}
\def\bd{\begin{displaymath}}
\def\ed{\end{displaymath}}

\usepackage{color}
\graphicspath{{.}{./Figures/}}
\begin{document}
%=====================================================
\title
{Scaling and Universality at Noisy Quench Dynamical Quantum Phase Transitions}

%=====================================================

\author{Saeid Ansari}
\email[]{ansari@bzte.ac.ir, saeid.ansari@gmail.com}
\affiliation{Department of Engineering Sciences and Physics, Buein Zahra Technical University, 34518-66391 Buein Zahra, Iran}

\author{R. Jafari}
\email[]{raadmehr.jafari@gmail.com}
%\affiliation{Department of Physics, Institute for Advanced Studies in Basic Sciences (IASBS), Zanjan 45137-66731, Iran}
%\affiliation{School of Quantum Physics and Matter Science, Institute for Research in Fundamental Sciences (IPM), 19395-5531, Tehran, Iran}
\affiliation{Physics Department and Research Center OPTIMAS, University of Kaiserslautern, 67663 Kaiserslautern, Germany}

\author{Alireza  Akbari}
\email[]{alireza@bimsa.cn}
\affiliation{Beijing Institute of Mathematical Sciences and Applications (BIMSA), Huairou District, Beijing 101408, China}

\author{Mehdi Abdi}
\email[]{mehabdi@gmail.com}
\affiliation{%Wilczek Quantum Center, 
School of Physics and Astronomy, Shanghai Jiao Tong University, 200240 Shanghai, China}

\date{\today}

\begin{abstract}
Dynamical quantum phase transitions (DQPTs) have been studied in the extended $XY$ model under both noiseless and noisy linear driven staggered field cases. 
In the time-independent staggered field case, the model exhibits a single critical point where the transition occurs from the spin-liquid phase to the antiferromagnetic phase.
In the noiseless ramp case, unlike the transverse field $XY$ model where DQPT always occurs for a quench crossing the single critical point, there is a critical sweep velocity above which the kinks corresponding to a DQPT are completely removed.
Furthermore, in this case there are only two \textit{critical modes} whose excitation probability is one-half.
In the presence of a Gaussian white noise, we find that this critical sweep velocity decreases by increasing the noise strength, and scales linearly with the square of  the noise intensity.
A surprising result occurs when the noise intensity and sweep velocity are about the same order of magnitude, the number of \textit{critical modes} is significantly increased, signalling a region with multiple critical modes.
Furthermore, our findings indicate that the scaling of the dynamical free energy near the DQPTs time is the same for both noiseless and noisy ramp quenches.
\end{abstract}

\maketitle
%-----------------------------------------------------------------------------
\section{Introduction}
In recent decades, there has been remarkable interest in studying the out-of-equilibrium physics of low-dimensional quantum systems~\cite{Polkovnikov2011,Cazalilla2010,Cazalilla2011,Bernard2016,Calabrese2016,Gogolin2016,Abanin2019,Mitra2018}. This renewed interest began with experimental advances in ultra-cold atoms in optical lattices~\cite{jotzu2014experimental,daley2012measuring,schreiber2015observation,choi2016exploring,flaschner2018observation}, which enabled the preparation and control of non-equilibrium states with unprecedented precision and stability~\cite{BlochRevModPhys2008}.
Following this, trapped ions~\cite{jurcevic2017direct,martinez2016real,neyenhuis2017observation,smith2016many}, nitrogen-vacancy centers in diamond~\cite{yang2019floquet}, superconducting qubit systems~\cite{guo2019observation}, and quantum walks in photonic systems~\cite{wang2019simulating,xu2020measuring} have further expanded the experimental framework for studying a wide range of non-equilibrium systems. These advances have also driven significant progress in theoretical physics.

Recently, a new theoretical framework called dynamical quantum phase transitions (DQPTs)~\cite{Heyl2013,Heyl2018} has been introduced for nonequilibrium quantum systems, which serve as a nonequilibrium counterpart to equilibrium phase transitions.
The concept of DQPT arises from the analogy between the equilibrium partition function of a system and the Loschmidt amplitude, which measures the overlap between an initial state and its time-evolved state~\cite{Heyl2013,Heyl2018,Jafari2019,Jafari2017,Divakaran2013,Guo2020,Najafi2018,Najafi2019,Yan2020,Zache2019,
Mukherjee2019,Wang2017,Zhang2016,Zhang2016b,Serbyn2017,Jafari2016}.
As equilibrium phase transitions are signaled by non-analyticities in thermal free energy, DQPTs are revealed through non-analytical behavior in dynamical free energy~\cite{andraschko2014dynamical,vajna2015topological,Karrasch2013,vajna2014disentangling,jafari2019dynamical,Mondal2022,Mendoza2022,Sedlmayr2018,Sedlmayr2018b,Khatun2019, Ding2020,Nicola2021,Verga2023,Rossi2022,Khan2023anomalous,Jad2021,Jad2024,Jad2023,Jad2021b,Vajna2014,Porta2020}, with real-time serving as the control parameter~\cite{Jafari2019,Jafari2017,Najafi2019,Sadrzadeh2021, Wong2022, Rylands2021,Abdi2019}.
DQPTs display phase transitions between dynamically emerging quantum phases during the nonequilibrium coherent quantum time evolution under sudden or ramped quenches~\cite{Zhou2021,Vanhala2023,Mondal2023,Cao2020,Sedlmayr2023,Sedlmayr2022,Sedlmayr2020,Uhrich2020,Zeng2023,Stumper2022,Yu2021,Vijayan2023,Xue2023,Bhattacharjee2023,Leela2022,Puskarov2016, Mishra2020,Divakaran2016,Sharma2016b,Zamani2024,jafari2024dynamical,Bagharan2024,Sacramento2024,Haldar2020,Ye2024,khan2023,Das2024} or time-periodic modulation of the Hamiltonian~\cite{yang2019floquet,Zamani2020,kosior2018dynamical,Jafari2021,kosior2018dynamicalb,Naji2022,Jafari2022,Naji2022b}.
In addition, a dynamical topological order parameter (DTOP) has been proposed to capture DQPTs~\cite{budich2016dynamical}, analogous to order parameters in equilibrium quantum phase transitions. The DTOP reveals integer values as a function of time, and its unit magnitude jumps at the dynamical phase transition times manifest the topological distinctive features of DQPTs~\cite{budich2016dynamical,Bhattacharya}.

DQPT has been observed experimentally in several studies~\cite{flaschner2018observation,jurcevic2017direct,martinez2016real,guo2019observation,wang2019simulating,Nie2020, Tian2020}, confirming theoretical predictions.
These studies are associated with deterministic quantum evolution generated by ramping or sudden quenching of the Hamiltonian.
However, comparatively little attention has been devoted to the stochastic driving of thermally isolated systems with a noisy Hamiltonian. In any real experiment, the simulation of the desired time-dependent Hamiltonian is imperfect, and noisy fluctuations are inevitable.
In other words, noise is ubiquitous and indispensable in any physical system. For example, noise-induced heating can originate from amplitude fluctuations of the lasers forming the optical lattice~\cite{Zoller1981,Chen2010,Doria2011,Marino2012,Marino2014}.
From a theoretical point of view, evolution with a noisy Hamiltonian extends the physics of disorder problems to the time domain, analogous to the disorder in real space, which has been widely studied for time-independent Hamiltonians~\cite{Anderson}.
The progresses in nonequilibrium systems motivate the examination of the effects of disorder in time for time-dependent Hamiltonians. Such investigations are of great practical and fundamental interest.
Thus, understanding the effects of noise on Hamiltonian evolution is crucial for (i) accurately predicting experimental outcomes and (ii) designing advanced experimental setups that are resilient to noise~\cite{Pichler}.

In this work, we contribute to expanding the systematic understanding of DQPTs in the extended $XY$ model under the driven staggered stochastic field. To this end, we introduce Gaussian white noise into the linear time-dependent staggered field. In the case of a time-independent staggered field, the model features only a single critical point where the transition occurs from the spin-liquid phase to the antiferromagnetic phase.
The results reveal that, in contrast to the transverse field $XY$ model, where dynamical quantum phase transitions (DQPT) take place for a quench that crosses the single critical point, there exists a critical sweep velocity beyond which DQPT are entirely eliminated.
In other words, for a quench crossing a single critical point, DQPTs always occur if the starting or ending point is confined between two critical points. 
We show that the critical sweep velocity decreases in the presence of noise and scales linearly with the square of noise intensity regardless of the noise strength. Moreover, we find that the effect of noise becomes particularly significant when the noise intensity and sweep velocity are in the same order of magnitude.
Under this condition,  the region of DQPTs splits into areas with two- or multi-critical modes. Crucially, we observe that the multi-critical modes region only appears in the presence of noise and does not appear for a noiseless system. We also demonstrate that the sweep velocity below which the system enters the multi-critical modes region scales linearly with the square of noise intensity for both weak and strong noise.

The paper is organized as follows: In Sec. \ref{RQDPT}, we discuss the dynamical free energy and the dynamical topological order parameter for two-band Hamiltonians. In Sec. \ref{model}, we present the model, its exact solution, and noiseless DQPTs. Section \ref{NRQ} is dedicated to the numerical simulation of the model based on the exact master equation in the presence of noise. Finally, Sec. \ref{Summary} contains concluding remarks.

%
%%%%%%%%%%%%%%%%%%%%%%%%%%%%  Fig. 1 %%%%%%%%%%%%%%%%%%%%%%%%%%%%%%%%
\begin{figure}[]
\centerline{\includegraphics[width=0.9\columnwidth]{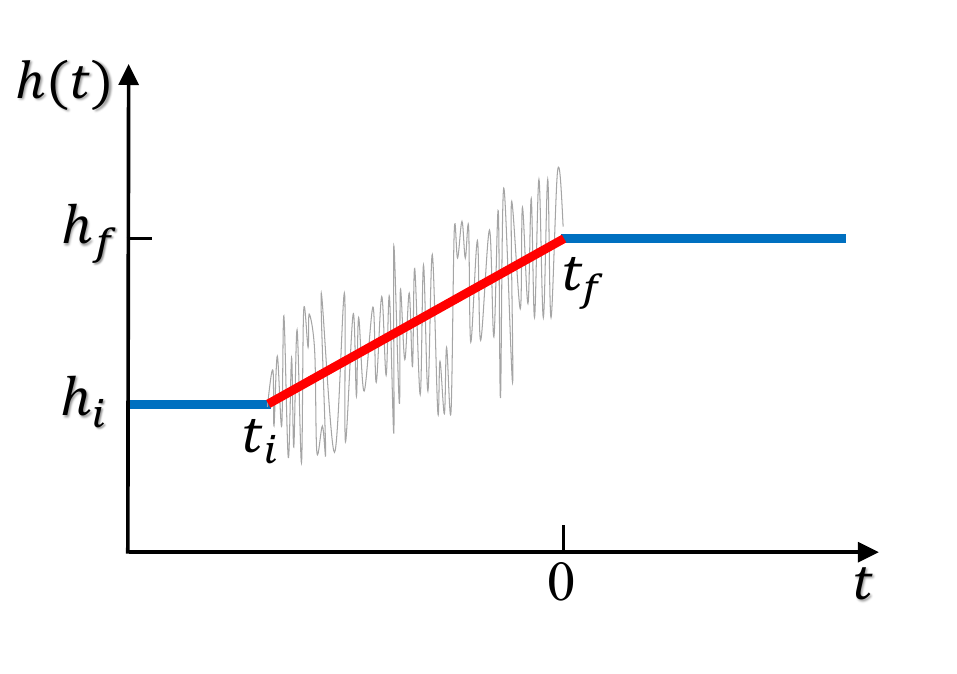}}
\caption{Schematically illustration of a linear ramp quench with noise fluctuation.
Quench starts at $t_i<0$, where magnetic field $h(t)$ set to $h_i$, and
ends up at $t_f\longrightarrow 0^{-}$, with $h(t_f)=h_f$.}
\label{fig1}
\end{figure}
%%%%%%%%%%%%%%%%%%%%%%%%%%%%%%%%%%%%%%%%%%%%%%%%%%%%%%%%%%%%%%%%%%%
%
%

%%%%%%%%%%%%%%%%%%%%%%%%%%%%%%%%%%%%%%%%%%%%%%%%%%%%%%%%%%%%%%%%%%%%%%%%%%%%%%%%%%%%%%%%
\section{Ramp Quench Dynamical Phase Transition in an Integrable Model\label{RQDPT}}
%%%%%%%%%%%%%%%%%%%%%%%%%%%%%%%%%%%%%%%%%%%%%%%%%%%%%%%%%%%%%%%%%%%%%%%%%%%%%%%%%%%%%%%%
%
%***************************************************************************************
\subsection{Dynamical free energy and Loschmidt overlap}
In this section, we describe ramp quench DQPTs, following the methodology introduced in Refs.~\cite{Divakaran2016,Sharma2016b,Zamani2024,Bagharan2024}. We focus on an integrable system that can be reduced to a two-level Hamiltonian $H_k(h)$ for each momenta mode $k$ with a tunable parameter $h$, i.e., $H(h)=\sum_k H_k(h)$.
In the ramped quench protocol, we assume that the parameter $h$ is quenched from an initial value $h_i$ at $t_i$ following the linear quench protocol
$h(t)=vt$, to a final value $h_f$ at time $t_f\rightarrow 0^{-}$ (Fig.~\ref{fig1}) in the manner that the system passes through the gap closing (critical) point $h=h_c$ during the ramp.
We assume that the system is initially prepared in the pre-quench Hamiltonian $H^i=\sum_k H_k(h_i)$ ground state $|g^i\rangle=\prod_k |g_k(h_i)\rangle$ at time $t_i$.
Since the system crosses the critical point the adiabatic evolution condition breaks down regardless of the quench speed $v$, which leads to a non-adiabatic transition. Therefore the final state $|\psi^f\rangle=\prod_k|\psi_k(h_f)\rangle$ is not generally the ground state of the post-quench Hamiltonian $H^f=\sum_k H_k(h_f)$.

Consequently, the final state for each $k$ mode is expressed as a linear combination of the ground and excited states
$|\psi^f_k\rangle=v_k|g^f_k\rangle+u_k|e^f_k\rangle$,
with the corresponding energy eigenvalues $\epsilon^{(g)}_k(h_f)=-\epsilon^f_k$ and $\epsilon^{(e)}_k(h_f)=\epsilon^f_k$, and the normalization condition $|u_k|^2+|v_k|^2=1$. 
As a parameter of interest that we will carefully study in this work the non-adiabatic transition probability, which represents the likelihood that the system ends up in the excited state after the quench, is given by $p_k=|u_k|^2=|\langle e^f_k|g^i_k\rangle|^2$.
For $t>t_f$ the dynamics of the system is characterized by the Loschmidt overlap ${\cal L}(t)=\prod_k {\cal L}_k(t)$  which is defined as
%
%-------------------------------  Eq1.  -----------------------------------------------
\bea
\label{eq1}
{\cal L}_k(t)&= \langle\psi^f_k|e^{-iH^f_kt}|\psi^f_k\rangle
=|v_k|^2 e^{-i\epsilon^f_{k}t}+|u_k|^2 e^{i\epsilon^f_{k}t}, 
\eea
%-------------------------------------------------------------------------------------
%
and corresponding dynamical free energy $g(t)=\frac{1}{N}\sum_k g_k(t)$ with $g_k(t)=-\ln|{\cal L}_k(t)|^2$, where $N$ is the size of the system. 
Summing over the contributions from all momentum modes $k$ and converting the summation to an integral in the thermodynamic limit, one obtains~\cite{Divakaran2016,Sharma2016b,Pollmann2013} 
%
%-------------------------------  Eq2.  -----------------------------------------------
\begin{align}\label{eq2}
%g_k(t)&=-\frac{1}{2\pi} \int_{0}^{\pi} dk \ln \left( |v_k|^2 e^{-i\epsilon^f_{k,g}t}+|u_k|^2 e^{-i\epsilon^f_{k,e}t}\right) \no\\
g(t)&=-\frac{1}{2\pi} \int_{0}^{\pi} dk \ln \left( 1-4p_k(1-p_k)\sin^2(\epsilon^f_kt)\right),
\end{align}
%-------------------------------------------------------------------------------------
%
where $t$ is measured from the instant the final state is reached at the end of the ramped quench $t_f$ (Fig.~\ref{fig1}).
The dynamical free energy $g(t)$ shows non-analytical behavior at real times
%
%-------------------------------  Eq3.  -----------------------------------------------
\begin{align}
\label{eq3}
t^*_n=\frac{\pi}
%{\epsilon^f_{k^{\ast}e}-\epsilon^f_{k^{\ast}g}}
{
\epsilon^{(e)}_{k^\ast}(h_f)- \epsilon^{(g)}_{k^\ast}(h_f)}
(2n+1)=\frac{\pi}{\epsilon^f_{k^{\ast}}}\left( n+\frac{1}{2} \right),
\end{align}
%-------------------------------------------------------------------------------------
%
provided that there exists critical momenta modes $k=k^{\ast}$, for which $p_{k^{\ast}}{=}1/2$. %\\
In other words, a half excitation probability for any post-quench mode leads to the occurrence of a sequence of cusps in the rate function at the transition times determined by Eq.~\eqref{eq3}.

%
%***************************************************************************************
\subsection{Dynamical Topological order parameter}
%***************************************************************************************
%
The dynamical topological order parameter (DTOP) is introduced to capture the topological characteristics associated with DQPTs\cite{budich2016dynamical}.
The DTOP only assumes integer values, and any change of its value at the critical times reveals the topological nature of those transitions~\cite{budich2016dynamical, Bhattacharjee2018,sharma2014loschmidt,Sharma2016b}. This dynamical topological order parameter, $N_w(t)$,  is extracted from the ``gauge-invariant" Pancharatnam geometric phase associated with the Loschmidt amplitude~\cite{budich2016dynamical,Bhattacharya}, and it is defined as~\cite{budich2016dynamical}
%
%=======================================  Eq. 4 ===============================
\begin{align}\label{eq4}
N_w(t)=\frac{1}{2\pi}\oint^\pi_0 dk\frac{\partial\phi^G(k,t)}{\partial k}.
\end{align}
%==========================================================================
%
Here, $\phi^G(k,t)$ is called geometric phase and can be deduced from two other phases
as 
$$\phi^G(k,t)=\phi(k,t)-\phi^D(k,t).$$
The total phase represented by $\phi(k,t)$, is extracted from the polar representation of the Loschmidt overlap, i.e.,
${\cal L}=|{\cal L}|\exp(i\phi(k,t))$ and the corresponding dynamical phase is
given as 
$$\phi^D(k,t)=-\int_{0}^{\pi}dt' \langle\psi^f_k(t')|H^f_k(t')|\psi^f_k(t')\rangle.$$
For two level systems, the total and the dynamical phases are given as follows~\cite{Divakaran2016,Sharma2016b}
%
%=======================================  Eq. 5 ===============================
\begin{equation*}
\phi(k,t)=\tan^{-1}\left(\frac{-|u_k|^2\sin(2\epsilon^f_kt)}{|v_k|^2+|u_k|^2\cos(2\epsilon^f_kt)}\right),\no
\end{equation*}\label{eq5}
\begin{equation*}
\phi^D(k,t)=-2|u_k|^2\epsilon^f_kt.
\end{equation*}
%==========================================================================
%
Therefore, the geometric phase is given by
%
%=======================================  Eq. 6 ===============================
\begin{align}\label{eq6}
\phi^G(k,t)=\tan^{-1}\left(\frac{-|u_k|^2\sin(2\epsilon^f_kt)}{|v_k|^2+|u_k|^2\cos(2\epsilon^f_kt)}\right)+2|u_k|^2\epsilon^f_kt.
\end{align}
%-------------------------------------------------------------------------------------
%
In the following section, we introduce the exactly solvable extended $XY$ spin model to study the noiseless ramp
quench DQPT.

%
%%%%%%%%%%%%%%%%%%%%%%%  Fig. 2   %%%%%%%%%%%%%%%%%%%%%%%
\begin{figure*}[t]
\begin{minipage}{\linewidth}
\centerline{\includegraphics[width=\linewidth]{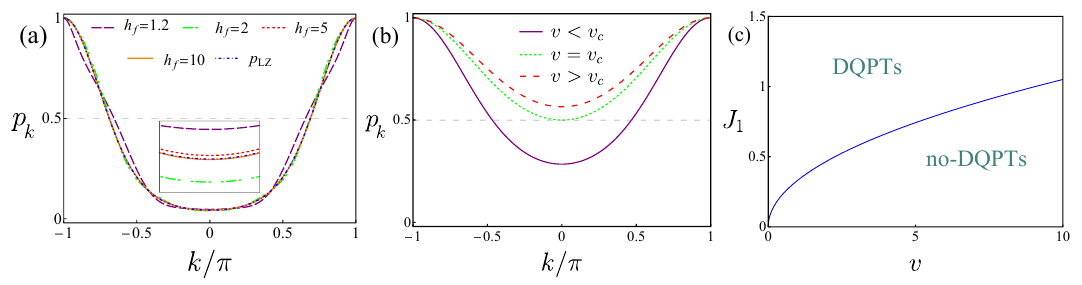}
}
\centering
\end{minipage}
%===============================================================
\caption{(a) Excitation probability following a linear ramp quench from the initial magnetic 
field $h_i=-10$ to the different values of $h_f$, for $v=2$.
Inset: the transition probability for $h_f=10$ is the same as Landau-Zener (LZ) 
transition probability (Eq. (\ref{eq:LZ})). 
(b) The excitation probability for $h_i=-10$ and $h_f=10$ for $v=5<v_c$, $v=v_c=9.06$ 
and $v=11>v_c$. (c)The dynamical phase diagram of
the model in the $v-J_1$ plane in the absence of the noise.}
\label{fig2}
\end{figure*}
%%%%%%%%%%%%%%%%%%%%%%%%%%%%%%%%%%%%%%%%%%%%%%%%%%%%%%%%%%%%
%

%%%%%%%%%%%%%%%%%%%%%%%%%%%%%%%%%%%%%%%%%%%%%%%%%%%%%%%%%%%%%%%%%%%%%%%%%%%%%%%%%%%%%%%%
\section{Extended $XY$ model: linear time dependent staggered magnetic field}\label{model}

The exactly solvable staggered field extended 
$XY$ model, in which the staggered magnetic field changes with time, is given as
%
%-------------------------------  Eq7.  -----------------------------------------------
\begin{align}
\label{eq7}
{\cal H}(t)&= \sum_{n=1}^{N}\Big[J_{1} \left(S_n^x S_{n+1}^x + S_n^y S_{n+1}^y \right)
+(-1)^n h^{(s)}(t) S_n^z\no 
\;
\\
&-(-1)^n J_{2} \left(S_n^x S_{n+1}^z S_{n+2}^x + S_n^y S_{n+1}^z S_{n+2}^y \right)\Big],
\end{align}
%-------------------------------------------------------------------------------------
%
where $N$ is the size of the system, $J_1$ is the coupling constant of the $XY$ interaction between nearest-neighbors~\cite{Jafari2008}, and $J_2$ denotes the strength of the staggered cluster (three-spin) interaction~\cite{Titvinidze}. It should be noted that, $J_1$, $J_2$ and $h^{(s)}$ have units of energy.
Here, we impose periodic boundary conditions, and $S_n^\alpha$ are the spin-1/2 operators at the $n$th site, i.e., $S_n^\alpha = \frac{1}{2} \sigma^{\alpha}_{n}$, where $\sigma^{\alpha = x, y, z}$ are Pauli matrices.
Without loss of generality we assume $J_1,J_2>0$.
As a preliminary, we investigate DQPT in the case of a noiseless linear ramp with sweep rate $v$ (in unit of energy per time), where 
$$h^{(s)}_{0}(t) = h^{(s)}_f + vt,$$
(the subindex in $h^{(s)}_{0}(t)$ is introduced to denote noise-free driving) starting from the initial value $h^{(s)}_{i}\equiv h_{i}$ at $t_i < 0$ and reaching the final value $h^{(s)}_{f}\equiv h_{f}$ when $t_f \rightarrow 0^{-}$.

The Hamiltonian in Eq.~\eqref{eq7} can be exactly diagonalized through the Jordan-Wigner transformation~\cite{Fabrizio1996,Suzuki1971a,SUZUKI1971b,Jafari2012}
%
%-------------------------------  Eq8.  -----------------------------------------------
\begin{align}\label{eq8}
S^{+}_{n}&= S^{x}_{n} + {\it i}S^{y}_{n}=\prod_{m=1}^{n-1}(1-2c_{m}^{\dagger}c_{m})c_{n}^{\dagger},\no \\
S^{-}_{n}&= S^{x}_{n} - {\it i}S^{y}_{n} = \prod_{m=1}^{n-1}c_{n}(1-2c_{m}^{\dagger}c_{m}),\\
S^{z}_{n}&= c_{n}^{\dagger}c_{n} -\frac{1}{2}, \no
\end{align}
%-------------------------------------------------------------------------------------
%
which transforms spin operators into spinless fermion counterparts  $c_{n}$, and $c^{\dagger}_{n}$.
The essential step is to define independent fermions at site $n$,
$ c_{n-1/2}^{A} \equiv c_{2n-1}$, $c_{n}^{B} \equiv c_{2n} $.
This effectively divides the chain into diatomic elementary cells. By introducing the Nambu spinor
$\Gamma^{\dagger}_k=(c_{k}^{\dagger B},~c_{k}^{\dagger A})$,
the Fourier transformed Hamiltonian can be expressed as a sum of independent terms acting in the two-dimensional Hilbert space generated by~$k$
%
%-------------------------------  Eq.11  -----------------------------------------------
\bea
\bl
\label{eq11}
{\cal H}(t)=\sum_{k}\Gamma_{k}^{\dagger}H^{(0)}_{k}(t)\Gamma_{k}
;~~
H^{(0)}_{k}(t)=2\left(
\begin{array}{cc}
-h_{z} & h_{xy} \\
h_{xy} & h_{z}\\
\end{array}
\right)
\el
\eea
%-------------------------------------------------------------------------------------
%
with $h_{xy}(k,t)=J_{1}\cos(k/2)$ and $h_{z}(k,t)=\frac{J_{2}}{2}\cos(k)+h^{(s)}_0(t)$,
where the wave number is $k=(2p-1)\pi/N$, $p=-N/4+1,-N/4+2, \cdots,  N/4$.\\

The eigenvalues and eigenvectors of the time-independent ($h^{(s)}_0(t)=h_0$) noninteracting Hamiltonian $\mathbb{H}_{k}(0)$ are
%
%-------------------------------  Eq.13  -----------------------------------------------
\begin{align}
\label{eq13}
\varepsilon^{\pm}_{k}=&\pm\varepsilon_k =\pm 2\sqrt{(h_{xy}(k))^{2}+(h_{z}(k))^{2}}, \no\\
|\chi^{+}_{k}\rangle=&\frac{1}{\sqrt{N_{k}}} \Big[h_{xy}|+\rangle+(h_{z}(k)-\varepsilon_{k})|-\rangle\Big],\\
\no
|\chi^{-}_{k}\rangle=&\frac{1}{\sqrt{N_{k}}} \Big[(h_{z}(k)-\varepsilon_{k})|+\rangle-h_{xy}|-\rangle\Big],
\end{align}
%------------------------------------------------------------------------------------
%
%
where $N_k=\sqrt{h_{xy}^2+(h_z(k) -\varepsilon_k)^2}$ is the normalization factor.
The energy gap between ground state and excited state, $\Delta_{k}=|\varepsilon^{+}_{k}-\varepsilon^{-}_{k}|$,
vanishes at Brillouin zone boundary $k_{c}=\pm\pi$, for $h^{(s)}_{c}=J_{2}/2$. Therefore, Quantum phase
transition occurs at critical point $h^{(s)}_{c}=J_{2}/2$, where the system transforms from the spin liquid phase
($h^{(s)}<J_{2}/2$) to the anti-ferromagnetic phase. 

The time dependent Schr\"{o}dinger equation of Hamiltonian in Eq. (\ref{eq11}) can be mapped to
the time dependent Schr\"{o}dinger equation of Landau-Zener (LZ) problem (see appendix \ref{AppendixA}),
which is exactly solvable~\cite{Vitanov1999,Vitanov1996,Suzuki2005,De_Grandi_2010}.
Consequently, if the system is prepared in its ground state at $t_i$ ($h^{(s)}_i\ll h^{(s)}_c$), the probability of finding the $k$th mode in the exited state (upper level) at finite time $t$ is given by non-adiabatic transition probability as~\cite{Vitanov1999} (see Appendix~\ref{AppendixA})
%
%-------------------------------  Eq.14  -----------------------------------------------
\bea
\bl
\label{eq14}
  p_k(\tau) 
  &
  =  e^{\frac{-\pi\Delta^2}{2v}}
  \Big|D_{i\frac{\Delta^2}{2v}}
  \left(2\sqrt{v}
  e^{\frac{i3\pi}{4}}
  \tau \right)\cos\theta(\tau) 
  \\
  &
    -\frac{\Delta}{\sqrt{v}}
    e^{\frac{-i\pi}{4}}
    D_{-1+i \frac{\Delta^2}{2v}}
    \left(2\sqrt{v}
    e^{\frac{i3\pi}{4}}
    \tau\right)\sin\theta(\tau)
    \Big|^2,
\el
\eea
%------------------------------------------------------------------------------------
%
where 
$\tau=t+\frac{J_2}{2v}\cos(k)$, and $\tan2\theta(\tau)=h_z/h_{xy}$, 
and $D_\nu(z)$ is the parabolic cylinder function~\cite{szego1954,Abramowitz1988}.
It should be mentioned that in the case $h^{(s)}_f \gg h^{(s)}_c$ the probability of excitations $p_k$ is given by the LZ transition probability as 
%
%-------------------------------  Eq.16  -----------------------------------------------
\begin{equation}
\label{eq:LZ}
p_{LZ}=
p_k(\infty)=
\exp
\Big(\frac{-2\pi J_1^2
\cos^2(\frac{k}{2})}{v}
\Big).
\end{equation}
%------------------------------------------------------------------------------------
%
For a ramp quench crossing the critical point $h^{(s)}_c$, the excitation probability $p_k$ after the quench largely depends on the value of $k$.

Indeed when the system is driven across the critical point, it undergoes nonadiabatic evolution and occupies the excited state due to the closing of the gap. 
Consequently, as illustrated in Fig.~\ref{fig2}(a), the transition probability becomes maximum at the gap-closing modes, i.e., $p_{k=\pm\pi} = 1$.
However, away from the gap-closing modes, the system evolves adiabatically for small sweep velocities due to the non-zero energy gap, which results in a small transition probability, see Fig.~\ref{fig2}(a).
Since the transition probability is a continuous function of $k$ in the thermodynamic limit, the condition for occurrence of the DQPT is to have $\min\{p_{k=0}\}\leq 1/2$. In that case, there exist critical mode(s) $k^{\ast}$ at which $p_{k^{\ast}} = 1/2$.
In other words, DQPT occurs if
%
%-------------------------------  Eq.17  -----------------------------------------------
\begin{equation}\label{eq17}
\min\{p_{k}\}=p_{k=0}=\exp{
\Big(
\frac{-2\pi J_{1}^{2}}{v}}
\Big)
\leq 1/2,
\end{equation}
%------------------------------------------------------------------------------------
%
which only holds for $v\leq v_{c}$ with $v_{c}=2\pi J_{1}^{2}/\ln(2)$.
Hence, there is a critical sweep velocity $v_{c}$ above which DQPTs are wiped out~\cite{Divakaran2016,Sharma2016b,Zamani2024,jafari2024dynamical,Bagharan2024}, and thus, DQPT does not occur after a sudden quench.
It is note worthy to remind that this is in contrast to the transverse field $XY$ model, where DQPTs always occur for a quench crosses a single critical point \cite{Divakaran2016,Sharma2016b,jafari2024dynamical}. 
This dynamical behavior in the extended $XY$ model is analogous to that of the transverse field $XY$ model when a quench crosses two critical points.
These findings imply that, for a quench crossing a single critical point, DQPT happens if the starting or ending point of the quench field is confined between two critical points. Otherwise, DQPT occur only for sweep velocities lower than the critical sweep velocity.

The transition probability is plotted in Fig.~\ref{fig2}(b) for $J_1 = 1$ and various values of sweep velocity. 
As predicted, $p_{k=\pm\pi} = 1$, and the minimum of $p_k$, which occurs at $k = 0$, is less than $1/2$ for $v < v_c$,
and DQPTs are completely wiped out for $v > v_c$. 
When $v < v_c$ this results two critical modes $k^{\ast}_{\alpha}$ and $k^{\ast}_{\beta}= -k^{\ast}_{\alpha}$, with $p_{k^{\ast}_{\alpha,\beta}} = 1/2$, 
and a sequence of DQPTs.
In Fig.~\ref{fig2}(c), we plot the dynamical phase diagram of the model in $v$-$J_1$ plane. The region marked ``DQPTs" supports periodic sequences of DQPTs, while in the ``no-DQPTs" region, DQPTs disappear entirely. 

%
%
%%%%%%%%%%%%%%%%%%%%%%%  Fig. 3   %%%%%%%%%%%%%%%%%%%%%%%
\begin{figure*}[t]
\begin{minipage}{\linewidth}
\centerline{\includegraphics[width=\linewidth]{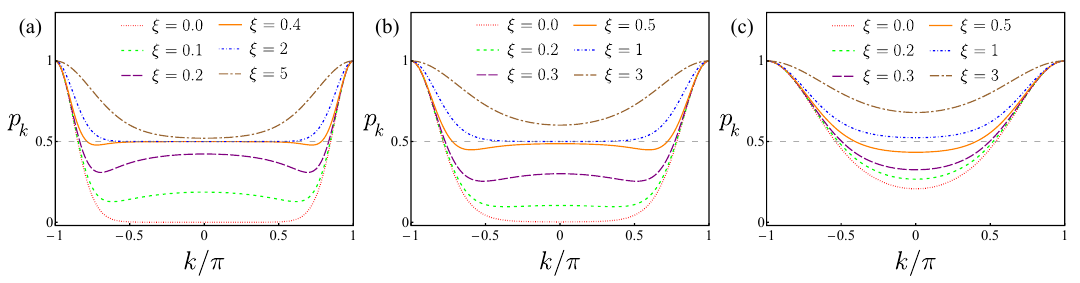}
}
\centering
\end{minipage}
%===============================================================
\caption{Probability of excitations for various noise's strength
for sweep velocity (a) $v=0.5$, (b) $v=1$, and (c) $v=4$.}
\label{fig3}
\end{figure*}
%%%%%%%%%%%%%%%%%%%%%%%%%%%%%%%%%%%%%%%%%%%%%%%%%%%%%%%%%%%%%%%%
%

%%%%%%%%%%%%%%%%%%%%%%%%%%%%%%%%%%%%%%%%%%%%%%%%%%%%%%%%%%%%%%%%%%%%%%%%%%%%%%%%%%%%%%%%
\section{Noisy Quench}\label{NRQ}
%%%%%%%%%%%%%%%%%%%%%%%%%%%%%%%%%%%%%%%%%%%%%%%%%%%%%%%%%%%%%%%%%%%%%%%%%%%%%%%%%%%%%%%%
%
Understanding the precise influence of noise is crucial in both classical and quantum regimes, since it is an inescapable and indispensable phenomenon in any physical system~\cite{Rahmani2016,Bando2020,Chenu2017,jafari2024dynamical,Bagharan2024,Asadian2025,Sadeghizade2025}.
Noise introduces time domain randomness into the system leading to disruption of coherent evolution.
In this section, we study the dynamical quantum phase transition in the extended $XY$ model, following the ramp quench with time dependent fluctuations.
To this aim, we add the uncorrelated Gaussian noise to the time dependent staggered magnetic field, i.e.,  
$$h^{(s)}(t)=h^{(s)}_f+vt+R(t),$$
where $R(t)$ is a random fluctuation with energy unit confined to the ramp interval $[t_i,t_f=0[$, with vanishing mean, $\langle R({t})\rangle=0$.
We use white noise with Gaussian two-time correlations 
$$\langle R(t)R(t')\rangle=\xi^2 \delta (t-t'),$$
where $\xi$, characterizes strength of the noise, and notice that $\xi^2$ has units of squared energy times time.

In order to specify the transition probability in the presence of noise, we solve numerically the exact master equation~\cite{Zamani2024,jafari2024dynamical,Bagharan2024,luczka1991quantum,Budini2000,Filho2017,Kiely2021} for
the averaged density matrix $\rho_{k}(t)$ of the Hamiltonian which includes noise 
$$H^{(\xi)}_k=H^{(0)}_{k}(t)+R(t)H_1,$$ 
where  $H^{(0)}_{k}(t)$ denotes noiseless Hamiltonian and $R(t)H_1$ indicates the added noise with $H_1=-2\sigma^z$ which appears during the ramp quench where $t\in [t_i,t_f=0[$. 
In the presence of uncorrelated Gaussian noise, the precise form of the master equation is provided as (Appendix \ref{AppendixB})
%
%-------------------------------  Eq.18  -----------------------------------------------
\begin{eqnarray}\label{eq18}
\bl\no
\frac{d}{dt}\rho_{k}(t)=-i\Big[H^{(0)}_{k}(t),\rho_{k}(t)\Big]-\frac{\xi^2}{2}\Big[H_1,[H_1,\rho_{k}(t)]\Big].
 \el
 \\
\end{eqnarray}
%--------------------------------------------------------------------------------------
%
which we take $\hbar=1$.
We numerically solve for the master equation, and employ the following equation to obtain the  transition probability in the presence of noise
%
%-------------------------------  Eq.19  -----------------------------------------------
\begin{eqnarray}
p_{k}= \langle \chi^+_{k}(t_f)|\rho_{k}(t_f)|\chi^+_{k}(t_f)\rangle.
\end{eqnarray}
%--------------------------------------------------------------------------------------
%
%
Consequently, the competition between noise-driven excitations and the near-adiabatic dynamics of the system's gapped fermionic modes characterizes the dynamical phase diagram of the model.

%
%===================================================
\begin{figure*}[t]
\includegraphics[width=1 \linewidth]{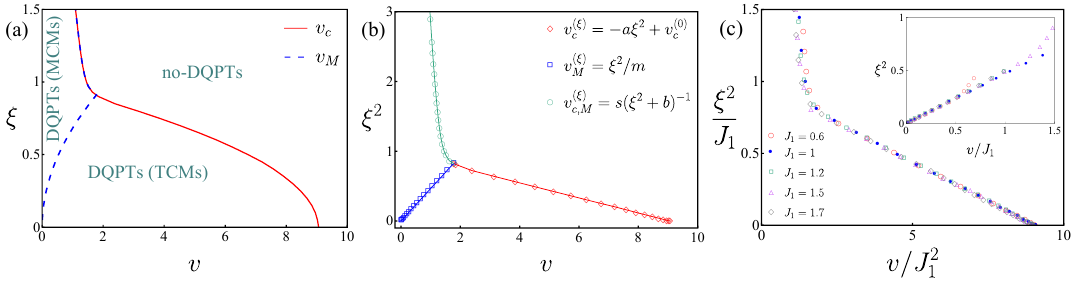}
\caption{(a) The phase diagram in the $v-\xi$ plane splitting
into DQPTs and no-DQPT regions. DQPTs region is classified into two sectors:
multi-critical modes (MCMs) (blue dashed line) and two critical
mode (TCM) (red solid line). (b) the critical sweep velocity $v_c$
(red solid line in Fig.~\ref{fig4}(a)) and the sweep velocity
below which the system enters to the MCMs region $v_M$ (blue dashed
line in Fig.~\ref{fig4}(a)) in $v-\xi^2$ plane. (c) A manifestation of scaling,
$v_c$  curves corresponding to different values of $J_1$ collapse to a single graph.
Inset: the scaling function of $v_M$.}
\label{fig4}
\end{figure*}
%===================================================
%

\subsection{Transition probability}
As previously stated, the dynamical quantum phase transitions for the quench that crosses the critical point only happen if the lowest transition probability, $p_{k=0}$, is less than $1/2$. This condition is fulfilled for $v<v_{c}=2\pi J_{1}^{2}/\ln(2)$.
This brings up the crucial question  about the robustness of DQPTs following a noisy quench.
Does a DQPT always happen, or will noise cause the critical sweep velocity to drop to zero? 
Anticipating the former is reasonable in light of our findings from crossing through a single equilibrium quantum critical point.

To simplify the analysis and without loss of generality we set the coupling constants $J_1=1$ and $J_2=2$ ($h_c=1$) and 
consider that $h^{(s)}_0(t)$ varying from $h^{(s)}_i=-10$ to $h^{(s)}_f=10$.
In Fig.~\ref{fig3} the transition probability is plotted versus $k$ for different values of sweep velocity and the noise intensity to understand the impact of noise on the transition probability.
As it can be seen, when $\xi/v\ll1$, the main effect of noise is to shift the critical mode(s) $k^{\ast}$, yielding DQPTs.
By gradually increasing the noise intensity, the effect of noise takes a surprising turn for $\xi/v \sim \mathcal{O}(1)$. 
In such a case, the critical modes with $p_k=1/2$ cover a finite interval of momenta which results-in multi-critical modes (MCMs). The locking of the $p_k$ curve to $1/2$ over a finite range of momenta suggests that the noise acts like a high-temperature source, resulting in maximally mixed states unless the $k$-modes are too “light” (easily excited to the upper level by Kibble-Zurek Mechanism~\cite{Rahmani2016}).
Furthermore, one finds that the main effect of noise during quenching time is that the inequality $p_{k=0}<1/2$ can be satisfied only for sufficiently low noise amplitudes. In other words, for $\xi/v\gg1$, very strong noise induces nonadiabatic transitions of such high probability that a maximally mixed state ($p_k=1/2$) does not appear at the end of the quench, thus blocking
appearance of DQPTs even for $v<v_{c}$.
Therefore, the boundary between "DQPTs" and "no-DQPTs" regions is modified in the presence of noise.
In the following we prob the dynamical phase diagram of the model for different noise intensity to better understand the effects of
noise on DQPTs.

\subsection{Dynamical phase diagram and scaling of critical sweep velocity}
The phase diagram of the model in the presence of noise is plotted in Fig.~\ref{fig4}(a) in the $v$-$\xi$ plane for $J_1=1$.
The numerical results reveal that in the presence of noise, the critical sweep velocity ($v^{(\xi)}_c$) above which the DQPTs disappear decreases with increasing noise intensity $\xi$ (red solid line).
This results are in good agreement with our expectation that noise induces nonadiabatic transitions, and a maximally mixed state ($p_k=1/2$) does not occur at the quench field end, thus blocking the appearance of DQPTs. Thus, for $v<v^{(\xi)}_c$, there are two critical modes at which DQPTs occur. 

The remarkable outcomes, however, appear when noise is present and the critical modes lock the $p_k$ curve to the value $1/2$ over a limited span of momenta known as the MCMs region.
Consequently, as shown in Fig.~\ref{fig4}(a), the DQPTs region splits into two regions in the presence of noise: the multi-critical modes region and the two critical modes (TCMs) region.  The blue dashed line denotes the sweep velocity $v_M$ that separates these zones from one another, while the red line denotes the critical sweep velocity $v^{(\xi)}_c$ that separates them from the non-DQPT region.
As can be observed, the critical sweep velocity $v^{(\xi)}_c$ (red solid line) decreases as noise rises, whereas the sweep velocity $v_M$ (blue dashed line) 
increases.  According to the numerical results, with extremely strong  noise ($\xi>1$), the $v_M$ curve meets the critical sweep velocity $v^{(\xi)}_c$ curve, and the two curves then combine to form a single curve.
\\

%
%%%%%%%%%%%%%%%%%%%%%%%  Fig. 5   %%%%%%%%%%%%%%%%%%%%%%%
\begin{figure*}[t]
\begin{minipage}{\linewidth}
\centerline{\includegraphics[width=\linewidth]{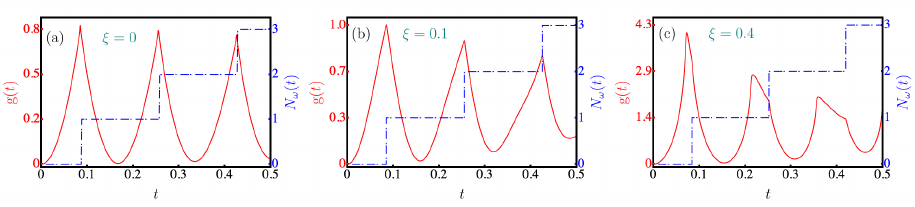}}
\centering
\end{minipage}
\caption{The dynamical free energy $g(t)$ (solid red line)
and associated DTOP, $Nw(t)$, (dashed blue line) for a quench from
$h_i=-10$ to $h_f=10$ for the Hamiltonian parameters $J_2=2$ and
$J_1=1$, $v=1$ (a) for noiseless case $\xi=0$, (b)
for noise intensity $\xi=0.1$, and (c) $\xi=0.4$.}
\label{fig5}
\end{figure*}
%%%%%%%%%%%%%%%%%%%%%%%%%%%%%%%%%%%%%%%%%%%%%%%%%%%%%%%%%%
%
%
%%%%%%%%%%%%%%%%%%%%%%%  Fig. 6   %%%%%%%%%%%%%%%%%%%%%%%
\begin{figure*}[t]
\begin{minipage}{\linewidth}
\centerline{\includegraphics[width=\linewidth]{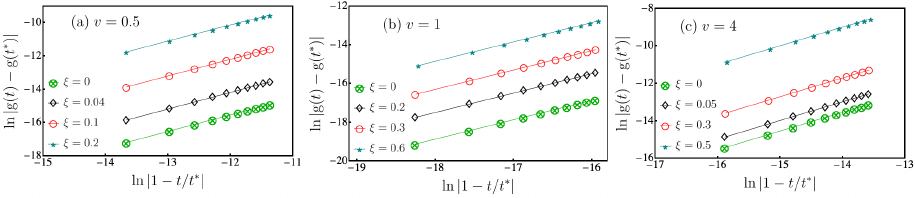}}
\centering
\end{minipage}
\caption{Scaling behavior of the dynamical free
energy in the vicinity if critical time $t^{\ast}$ in the
presence of various noise’s strength  for different values
of the sweep velocity (a) $v=0.5$, (b) $v=1$ and (c) $v=4$.}
\label{fig6}
\end{figure*}
%%%%%%%%%%%%%%%%%%%%%%%%%%%%%%%%%%%%%%%%%%%%%%%%%%%%%%%%%%%%%%%%%%%%%
%

%
%%%%%%%%%%%%%%%%%%%%%%%  Fig. 7   %%%%%%%%%%%%%%%%%%%%%%%
\begin{figure*}[t]
\begin{minipage}{\linewidth}
\centerline{\includegraphics[width=\linewidth]{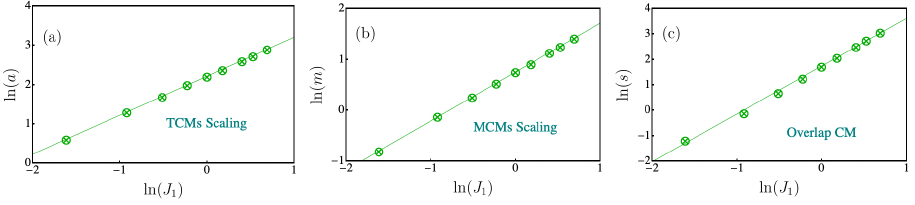}
}
\centering
\end{minipage}
%===============================================================
\caption{Scaling of the slope of linear lines
corresponding to Fig.~\ref{fig4}(b)  with $J_1$.}
\label{fig7}
\end{figure*}
%%%%%%%%%%%%%%%%%%%%%%%%%%%%%%%%%%%%%%%%%%%%%%%%%%%%%%%%%%%%%%%%%
%

%
%%%%%%%%%%%%%%%%%%%%%%%  Fig. 8   %%%%%%%%%%%%%%%%%%%%%%%
\begin{figure}[t]
\begin{minipage}{\linewidth}
\centerline{\includegraphics[width=\linewidth]{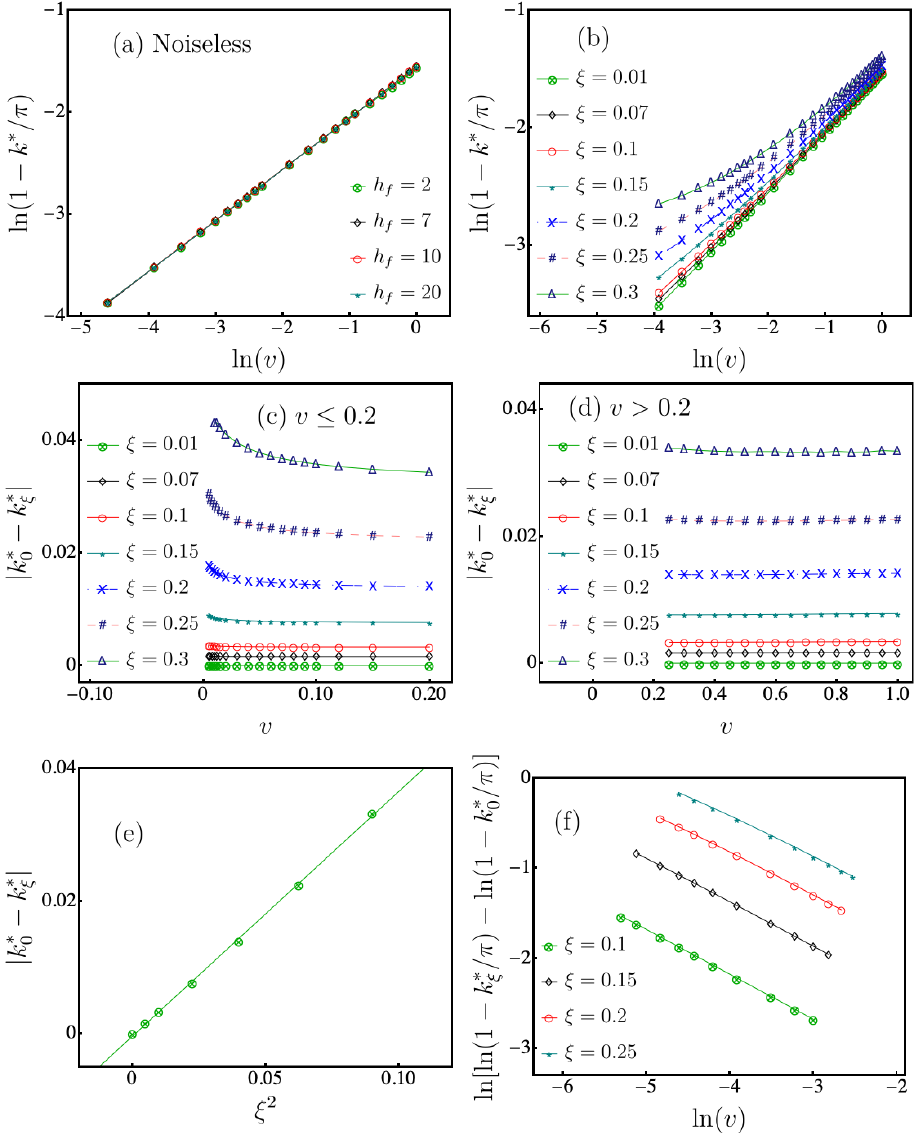}}
\centering
\end{minipage}
%\vspace{0.5cm}
%===============================================================
\caption{
(a) Scaling of critical mode $k^{\ast}$ versus
sweep velocity for noiseless case for various post-quench magnetic field ($h_f>h_c=1$) .(b) Critical mode $k^{\ast}$ versus
sweep velocity in the presence of noise. (c) and (d) Deviation of critical $k$ in noisy
model from noiseless one, against sweep velocity for $v\leq 0.2$ and $v>0.2$,  respectively. (e)  Deviation of critical $k$ in noisy
model from noiseless one, against noise strength for plot (d). (f) Similar plot with logarithmic values.}
\label{fig8}
\end{figure}
%%%%%%%%%%%%%%%%%%%%%%%%%%%%%%%%%%%%%%%%%%%%%%%%%%%%%%%%
%

An intriguing unresolved topic from a physical perspective is how noise affects the phase diagram: 
does the system exhibit any universality and scalability in the presence of noise? 
To study this, in Fig.~\ref{fig4}(b) we display the critical sweep velocities, $v^{(\xi)}_c$ and $v_M$, in the $v$-$\xi^2$ plane.
The results of our numerical analysis show that, for both weak ($\xi \sim \mathcal{O}(10^{-2})$) and strong ($\xi \sim \mathcal{O}(10^{-1})$) noise, the critical sweep velocity scales linearly with the square of noise intensity. In other words, $v^{(\xi)}_c=-a\;\xi^2+v^{(0)}_c$, where $v^{(0)}_c=2\pi J_{1}^{2}/\ln(2)$ represents the critical sweep velocity in the noiseless case. Furthermore, a similar linear scaling is also seen for $v_M$, which is defined as $v^{(\xi)}_M=\xi^2/m$, in the presence of both weak and strong noise.
The critical sweep velocity also follows the scaling $v^{(\xi)}_{c,M}=s\;(\xi^2+b)^{-1}$ for very strong noise ($\xi>1$) when both $v_c$ and $v_M$ merge.

A more detailed analysis shows that the slope of lines in Fig.~\ref{fig4}(b) also scales with
coupling $J_1$ (see Fig.~\ref{fig7} in Appendix \ref{AppendixC}). The numerical results reveal that the slops $a$, $m$ and $s$ scale in a power law manner with
$J_1$ with exponent $\beta$, i.e., $\{a,m,s\}\propto J_1^{\beta}$ with $\beta=-1\pm0.01$ for $\{a,m\}$ and $\beta=-2\pm0.01$ for $s$.
The numerical findings appear to indicate an extrapolation for a scaling behaviour of $v^{(\xi)}_c$, i.e., $v^{(\xi)}_c$ is invariant under the scaling transformations $\xi\rightarrow\xi/\sqrt{J_1}$ and $v^{(\xi)}_c\longrightarrow v^{(\xi)}_c/J_1^2$.
Additionally, $v_M\rightarrow v_M/J_1$ is the scaling function for multi-critical sweep velocity.
In this regard, Fig.~\ref{fig4}(c) illustrates the scaling of critical sweep velocity $v^{(\xi)}_c$ corresponding to various values of $J_1$, for both weak and strong noise. 
It illustrates how all curves collapse into a single graph under the scaling function.
The inset of Fig.~\ref{fig4}(c) also shows the scaling function for $v_M$ below which the system reveals multi-critical modes. 
These scaling functions represent the promised universality of DQPTs in the face of noise, which motivates us to look for the dynamical free energy's universality.

%
%%%%%%%%%%%%%%%%%%%%%%%%%%%%%%%%%%%%%%%%%%%%%%%%%%%%%%%%%%%%%%%%%%%%%%%%%%%%%%%%%%%%%%%%%%%%%%
\subsection{Scaling of dynamical free energy}
Finally, in this section we study the scaling of the rate function and its topological aspects in the presence of noise.
The dynamical free energy and corresponding dynamical topological order parameter are shown in Fig.~\ref{fig5}(a) for the noiseless case ($\xi=0$), Fig.~\ref{fig5}(b) for $\xi=0.1$, and Fig.~\ref{fig5}(c) for $\xi=0.4$, which corresponds to Fig.~\ref{fig3}(b).
For noiseless and weak noise cases the clear cusps in $g(t)$ and the quantization and jumps in the associated DTOP are clearly visible.
In Fig.~\ref{fig3}(b), the shift in the critical modes for $\xi=0.1$ is very tiny which results in very small changes in the dynamical quantum phase transition critical times $t^{\ast}$.
Nevertheless, for a strong noise $\xi=0.4$ the cusps in $g(t)$ are not clearly visible but by a careful examination the occurrence of cusps at the critical times can be observed. This is further supported by the readily noticeable jumps in the corresponding DTOP as an indicator of DQPTs, (Fig.~\ref{fig5}(c)). 
The extra picks in $g(t)$ in the case of strong noise are indeed smooth and result from the appearance of \textit{nearly} maximally mixed state at $k=0$, which is clearly visible in the transition probability for $\xi=0.4$ in Fig.~\ref{fig3}(b).
A note should be made that even though the system has two critical modes $k^{\ast}_\beta=-k^{\ast}_\alpha$ at which $p_k^{\ast}=1/2$, the symmetry of the eigenvalues ($\pm\varepsilon_k(h_f)$) with regard to $k=0$ causes the dynamical free energy to disclose a single critical time scale.

Similar to equilibrium phase transitions, one of the main challenges with the DQPT concept is to look for fundamental concepts like universality and scaling.
Having established appearance of DQPTs in the presence of noise, we now turn to address the question of scaling and universality at dynamical free energy. 
In a one-dimensional system, it has been shown that, for a sudden quench, the dynamical free energy indicates power-law scaling in the  vicinity of the critical time, i.e., $|g(t)-g(t^*)|\sim|(t-t^*)/t^*|^\nu$, with $\nu=1$~\cite{Heyl2015PRL}.
Moreover, the same scaling has been also confirmed for noiseless ramp quench in a one-dimensional system for all values of sweep velocity~\cite{Zamani2024}.
While for two-dimensional systems the scaling law is logarithmic for a sudden quench case~\cite{Heyl2015PRL}.
In a similar vein, we have probed the scaling behavior of the dynamical free energy close to the DQPT time scale $t^*$,
and the findings are represented in Fig.~\ref{fig6}. The numerical results shows that the dynamical free energy in the presence of noise scales in a power-law manner with exponent $\nu\simeq 1$, for all values of the sweep velocity and any noise intensity, similar to that of the noiseless case~\cite{Zamani2024}. This means that the universal scaling exponent does not change in the presence of a white noise.
It should be emphasize that in the sudden quench case the model does not show DQPTs.
It is important to note that we assumed the system's state after the quench to be the noise-averaged pure state, and the dynamical free energy is derived under this assumption; therefore, the reliability of our results hinges on this assumption.
%
%%%%%%%%%%%%%%%%%%%%%%%%%%%%%%%%%% Discussion  %%%%%%%%%%%%%%%%%%%%%%%%%%%%%%%%%%%%%%%%%%%%%

\section{Discussion}\label{Summary}
In this study, we have looked at the extended $XY$ model's dynamical quantum phase transitions when it is driven by the staggered stochastic field. The model displays just one critical point in the time-independent staggered field. Unlike the transverse fields $XY$ model, that DQPTs always occur for a sudden and ramp quench across a single critical point, the extended $XY$ model never exhibits DQPT under the sudden quench protocol and DQPT occurs in the noiseless and noisy ramp quench case only for slow driven staggered magnetic field.
In other words, our findings show that the extended $XY$ model for a quench that passes a single critical point exhibits the same dynamic behavior as the transverse field XY model for a quench that crosses two critical points.
From a wider point of view, this result imply that for a sudden quench that crosses a single critical point, the DQPTs appear only if either the starting or the ending point of the quench field is confined between two critical points. Nonetheless, even though the DQPTs is absent in the sudden quench case, there is a critical sweep velocity below which the DQPTs can occur for a ramped quench with finite speed.

In the noiseless ramp quench, the system contains two critical modes at which DQPTs occur if the sweep velocity is less than the critical sweep velocity.
In the presence of white noise, the critical sweep velocity above which the DQPTs are wiped out decreases by increasing the noise intensity and scales linearly with the square of noise intensity. Moreover, added noise to the staggered magnetic field creates the region with multi-critical modes.
According to the numerical simulation,  the sweep velocity ($v_M$) below which the system enters to the multi-critical modes region, also scales linearly with the square of noise intensity. In the weak and strong noise case, we have also found the scaling function under which the critical sweep velocity and $v_M$ for different values of the Hamiltonian's parameters collapse on the single curve.
In addition, the numerical results manifest that the dynamical free energy close to the DQPTs time shows power law scaling with the exponent $\nu\simeq1$ for all sweep velocities and any noise intensity which is the same as the noiseless ramp quench scaling features of the one dimensional systems~\cite{Zamani2024}.

It is important to note that our findings, are agreement with the previous work on transverse field Ising model with cluster interaction which reveals three critical points~\cite{Jafari2025c}.
At the end, it is worth noting that the critical modes in the noisy ramp quench scenario also exhibit scaling with sweep velocity, just as they do in the noiseless case,
where they exhibit linear scaling with square root of the sweep velocity. 
The details of scaling of the critical modes have been represented in Appendix \ref{AppendixD}.
%

%
%%%%%%%%%%%%%%%%%%%%%%%%%%%%%%%%%% Appendix  %%%%%%%%%%%%%%%%%%%%%%%%%%%%%%%%%%%%%%%%%%
\appendix

%%%%%%%%%%%%%%%%%%%%%    Appendix A          %%%%%%%%%%%%%
\section{Time dependent Schr\"{o}dinger in the diabatic basis}\label{AppendixA}
The time dependent Schr\"{o}dinger equation of the Hamiltonian given in Eq.~(\ref{eq11})
can be written as
%--------------- Eq App1 -----------------
\begin{eqnarray}
i\frac{d}{dt}
\left(
\begin{array}{c}
a_1(t) \\
a_2(t)
\end{array}
\right)=2\left(
\begin{array}{cc}
-h_{z} & h_{xy} \\
h_{xy} & h_{z}\\
\end{array}
\right)
\left(
\begin{array}{c}
a_1(t) \\
a_2(t)
\end{array}
\right),
\end{eqnarray}
with $h_{xy}(k,t)=J_{1}\cos(k/2)$ and $h_{z}(k,t)=\frac{J_{2}}{2}\cos(k)+h_{s}(t)$, 
where  $h_{s}(t)=h_f+vt$.
%\\
By defining new variables as given
\begin{equation}
 \begin{aligned}
 \tau= t+\frac{J_2}{2v}\cos(k),
 \qquad
\Delta=J_1\cos(k/2),
\end{aligned}
\end{equation}
the Hamiltonian
\bea
%\label{eq25}
H^{(0)}_{k}(t)&=&2\left(
\begin{array}{cc}
-h_{z} & h_{xy} \\
h_{xy} & h_{z}\\
\end{array}
\right)
\eea
can be mapped to Landau-Zener counterpart $\mathbb{H}_{\text{LZ}}(\tau)$
\bea
%\label{eq25}
\mathbb{H}_{\text{LZ}}(\tau)=2\left(
\begin{array}{cc}
v\tau & \Delta \\
\Delta &-v\tau\\
\end{array}
\right),
\eea
which then can be solved analytically~\cite{Vitanov1996,Vitanov1999}.
%%%%%%%%%%%%%%%%%%%%%%%%%%%%%%%%%%%%%%%%%%%%%%%%%%%%%%%%%%%%%%%%%%%%%%%%%%%%%%
%

%
%%%%%%%%%%%%%%%%%%%%%%%%%%%%%%%%   Appendix B    %%%%%%%%%%%%%%%%%%%%%%%%%%%%%%%%%
\section{Ensemble-averaged transition probabilities: Exact noise master equation}\label{AppendixB}

In the following we outline the procedure by which the noise exact master equation is calculated.
For transparency and ease of notation, we begin by considering a general time-dependent Hamiltonian \cite{Kiely2021},
%
%%%%%%%%%%%%%%%%%%%%%%%%%%%%%%%%%%%%%%%%%%%%%%%%%%%%%%%%%%%%%
\begin{equation} 
\label{Hamiltonian}
H(t)=H_{0}(t)+R(t)H_{1}(t),
\end{equation}
%%%%%%%%%%%%%%%%%%%%%%%%%%%%%%%%%%%%%%%%%%%%%%%%%%%%%%%%%%%%%%
%
where $H_{0}(t)$ is noise-free while $H_{1}(t)$ is ``noisy" with $R(t)$ a real function for a given realization of the noise. 
We have considered noise $R(t)$ with mean $\langle R(t)\rangle=0$, and correlation function defined as 
%
%%%%%%%%%%%%%%%%%%%%%%%%%%%%%%%%%%%%%%%%%%%%%%
\begin{equation}
\label{OU}
\langle R(t)R(t')\rangle=C(t,t')=C(t-t'),
\end{equation}
%%%%%%%%%%%%%%%%%%%%%%%%%%%%%%%%%%%%%%%%%%%%%%%%%%%%%%%%%%%%%%
%
With this setup we now derive a noise master equation for the averaged density matrix $\rho(t)$ of 
$H(t)$ \cite{luczka1991quantum,Kiely2021}. One starts by writing down the von Neumann equation
%
%%%%%%%%%%%%%%%%%%%%%%%%%%%%%%%%%%%%%%%%%%%%%%%%%%%%%%%%%%%%%
\begin{equation}
\label{Neumann}
\dot{\rho}_{R}(t)=-\frac{i}{\hbar}[H(t),\rho_{R}(t)],
\end{equation}
%%%%%%%%%%%%%%%%%%%%%%%%%%%%%%%%%%%%%%%%%%%%%%%%%%%%%%%%%%%%%
%
where
%%%%%%%%%%%%%%%%%%%%%%%%%%
\begin{equation} 
\label{quenchevolved}
\rho_{R}(t) = U^{\dagger}_{R}(t,t_i)\rho_{R}(t_i)U_{R}(t,t_i)
\end{equation}
%%%%%%%%%%%%%%%%%%%%%%%%%%
is the density matrix for a specific realization of the noise function $R(t)$. As follows from Eq. (\ref{Neumann}), the time-evolution from the noise-free initial condition $\rho_{R}(t_i)$ at time $t_i$ is carried through by $U_{R}(t,t_i) = {\cal T}\exp(-\frac{i}{\hbar}\int_{t_i}^t H (t^{\prime})\, dt^{\prime})$, with ${\cal T}$ the time-ordering operator.
Introducing $\rho(t) = \langle \rho_R(t) \rangle$ as the ensemble average over many noise realizations (all with a common noise-free initial condition), the averaged von Neumann equation (\ref{Neumann}) takes the form
%
%%%%%%%%%%%%%%%%%%%%%%%%%%%%%%%%%%%%%%%%%%%%%%%%%%%%%%%%%%%%%
\begin{equation}
\label{meanNeumann}
\dot{\rho}(t)=-\frac{i}{\hbar}[H_0(t),\rho(t)] - \frac{i}{\hbar}[H_1(t),\langle R(t) \rho_R(t)\rangle].
\end{equation}
%%%%%%%%%%%%%%%%%%%%%%%%%%%%%%%%%%%%%%%%%%%%%%%%%%%%%%%%%%%%%%%%%%%%%%%%
%
Applying a theorem by Novikov \cite{Novikov1965} one has 
%
%%%%%%%%%%%%%%%%%%%%%%%%%%%%%%%%%%%%%%%%%%%%%%%%%%%%%%%%%%%%%
\begin{equation}
\label{Novikov}
\langle R(t) \rho_R(t) \rangle = \langle R(t) \rangle \langle \rho_R(t) \rangle + \int_{t_i}^t \mbox{d}s \langle R(t) R(s) \rangle \langle \frac{\delta \rho_{R}}{\delta \eta}\rangle,
\quad
\end{equation}
%%%%%%%%%%%%%%%%%%%%%%%%%%%%%%%%%%%%%%%%%%%%%%%%%%%%%%%%%%%%%%%%%%%%%
%
with functional derivative
%
%%%%%%%%%%%%%%%%%%%%%%%%%%%%%%%%%%%%%%%%%%%%%%%%%%%%%%%%%%%%%
\begin{equation}
\label{func}
\frac{\delta \rho_{R}}{\delta R} = \frac{\partial\dot{\rho}_{R}}{\partial R} - \frac{\mbox{d}}{\mbox{d}t} \frac{\partial\dot{\rho}_{R}}{\partial\dot{R}}.
\end{equation}
%%%%%%%%%%%%%%%%%%%%%%%%%%%%%%%%%%%%%%%%%%%%%%%%%%%%%%%%%%%%%%%%%
%
Combining Eq. (\ref{func}) with (\ref{Hamiltonian}) and (\ref{Neumann}) gives
%
%%%%%%%%%%%%%%%%%%%%%%%%%%%%%%%%%%%%%%%%%%%%%%%%%%%%%%%%%%%%%
\begin{equation}
\label{func2}
\frac{\delta \rho_{R}}{\delta R} = -\frac{i}{\hbar}[H_1(t),R(t)].
\end{equation}
%%%%%%%%%%%%%%%%%%%%%%%%%%%%%%%%%%%%%%%%%%%%%%%%%%%%%%%%%%%%%%%
%
%
The master equation follows by inserting Eq. (\ref{Novikov}) into (\ref{meanNeumann}), using Eqs. (\ref{OU}) and (\ref{func2}),
%
%%%%%%%%%%%%%%%%%%%%%%%%%%%%%%%%%%%%%%%%%%%%%%%%%%%%%%%%%%%%%
\begin{equation}
\bl
\label{master}
\dot{\rho}(t)
=
&-\frac{i}{\hbar}[H_{0}(t),\rho(t)]\\
&- \frac{1}{\hbar^2}\Big[H_{1}(t),\int_{t_i}^{t}C(t-s)[H_{1}(t),\rho(s)]ds\Big].
\el
\end{equation}
%%%%%%%%%%%%%%%%%%%%%%%%%%%%%%%%%%%%%%%%%%%%%%%%%%%%%%%%%%%%%%
%
The first term on the right-hand side accounts for the unitary time evolution generated by the prescheduled noiseless Hamiltonian
$H_{0}(t)$ and the second term induces the dynamics noise with Hamiltonian $H_{1}(t)$.
The most common case is Gaussian white noise which is completely uncorrelated in time which is
defined by a correlation function $C(t-t')=\langle R(t)R(t')\rangle=\xi^2 \delta (t-t')$. Therefore, 
the master equation Eq. (\ref{master}) simplifies to \cite{luczka1991quantum,Kiely2021,Budini2000}
%
%%%%%%%%%%%%%%%%%%%%%%%%%%%%%%%%%%%%%%%%%%%%%%%%%%%%%%%%%%%%%
\begin{equation}
\bl
\label{master2}
\dot{\rho}(t)=-\frac{i}{\hbar}[H_{0}(t),\rho(t)]
- \frac{\xi^2}{2\hbar^2}\Big[H_{1}(t),[H_{1}(t),\rho(t)]\Big].
\quad
\el
\end{equation}
%%%%%%%%%%%%%%%%%%%%%%%%%%%%%%%%%%%%%%%%%%%%%%%%%%%%%%%%%%%%%%
%
It is worth mentioning that, if the energy scale is introduced in the Hamiltonian, in such a way that coupling constants 
in the Hamiltonian become dimensionless, the sweep velocity $v$, and square of noise intensity $\xi^2$
have unit of inverse time and time respectively and $R(t)$ is dimensionless random number \cite{Rahmani2016}.

%%%%%%%%%%%%%%%%%%%%%%%%%%%%%%%%  Appendix C  %%%%%%%%%%%%%%%%%%%%%%%%%%%%%%%%%%%%%%%%%
\section{Scaling of the slope}\label{AppendixC}
In Fig.\ref{fig7}, the scaling of the slopes of the linear lines from Fig.\ref{fig4}(b) is shown as a function of $J_1$. As observed, in all cases, the slope of the lines varies linearly with $J_1$.

%%%%%%%%%%%%%%%%%%%%%%%%%%%%%%%%  Appendix D  %%%%%%%%%%%%%%%%%%%%%%%%%%%%%%%%%%%%%%%%%
\section{Scaling of the critical modes in the presence of noise}\label{AppendixD}
In Fig.~\ref{fig8}, we display several scalings of the critical mode. As Fig.~\ref{fig8}(a) illustrates, in the noiseless model, dependence of the critical mode to the sweep velocity is independent of the post-quench value of magnetic field $h_f$ and their logarithms scales linearly for $|v|\leq 1$~\cite{Zamani2024}. For models with various coupling constant $J_1$ values, we observe shifted linear scaling. When noise is taken into account, Fig.~\ref{fig8}(b), the scaling linearity breaks down and, as the noise grows, the deviation increases. Furthermore, the deviation of critical mode in the presence of noise $k^{\ast}_\xi$, compared to the noiseless case $k^{\ast}_0$ is indicated in Fig.~\ref{fig8}(c). Here, we witness two different behaviors; When sweep velocity $v$ is very small,$|k^{\ast}_0-k^{\ast}_\xi|$ depends of the velocity (see Fig.~\ref{fig8}(c)), however as the sweep velocity increases, this deviation remains independent of velocity (see Fig.\ref{fig8}(d)).  Fig.\ref{fig8}(e) shows that in the later case, $|k^{\ast}_0-k^{\ast}_\xi|$ scales linearly by $\xi$-squared.
we gain additional insights in Fig.~\ref{fig8}(f) by considering logarithm of logarithmic differences of noisy $k^{\ast}$s relative to noiseless case and witness linear scaling with sweep velocity.

%%%%%%%%%%%%%%%%%%%%%%%%%%%%%%%%%% END Appendix  %%%%%%%%%%%%%%%%%%%%%%%%%%%%%%%%%%%%%%%%%%
%
%%%%%%%%%% Biblography %%%%%%%%%%%%%%%%%%%%
\bibliography{Ref_DQP_Noisy}
%-----------------------------------------------------------------------------
\end{document}